\documentclass[times,twocolumn,final,authoryear]{elsarticle}

\usepackage{jasr}
\usepackage{framed,multirow}

\usepackage{amssymb}
\usepackage{latexsym}
\usepackage{natbib}
\usepackage{amsmath}
\usepackage{gensymb}

\usepackage{url}
\usepackage{xcolor}
\definecolor{newcolor}{rgb}{.8,.349,.1}
\usepackage{graphicx}
\usepackage[citebordercolor=white]{hyperref}

\journal{Advances in Space Research}

\begin{document}

\verso{Jean-Marie Malherbe \textit{etal}}

\begin{frontmatter}

\title{The SLED project and the dynamics of coronal flux ropes}

\author[1]{Jean-Marie \snm{Malherbe}}
\ead{Jean-Marie.Malherbe@obspm.fr}
\author[1]{Pierre \snm{Mein}}
\ead{Pierre.Mein@club-internet.fr}
\author[2]{Fr\'{e}d\'{e}ric \snm{Say\`{e}de}}
\ead{Frederic.sayede@obspm.fr}
\author[3]{Pawel \snm{Rudawy}}
\ead{rudawy@astro.uni.wroc.pl}
\author[4]{Kenneth \snm{Phillips}}
\ead{kennethjhphillips@yahoo.com}
\author[5]{Francis \snm{Keenan}}
\ead{f.keenan@qub.ac.uk}
\author[6]{Jan \snm{Ryb\'{a}k}}
\ead{rybak@astro.sk}

\address[1]{Observatoire de Paris, LESIA, 92195 Meudon, France}
\address[2]{Observatoire de Paris, GEPI, 92195 Meudon, France}
\address[3]{Astronomical Institute, University of Wroc{\l}aw, Poland}
\address[4]{Earth Sciences Department, Natural History Museum, London SW75BD, United
              Kingdom}
\address[5]{Astrophysics Research Centre, School of Mathematics and Physics, Queen's University Belfast, United
              Kingdom}
\address[6]{Astronomical Institute, Slovak Academy of Sciences, 05960 Tatransk\'{a} Lomnica,
              Slovakia}

\received{1 May 2021} \finalform{-} \accepted{-} \availableonline{-}
\communicated{-}

\begin{abstract}
Investigations of the dynamics of the hot coronal plasma are crucial
for understanding various space weather phenomena and making
in-depth analyzes of the global heating of the solar corona. We
present here numerical simulations of observations of siphon flows
along loops (simple semi-circular flux ropes) to demonstrate the
capabilities of the Solar Line Emission Dopplerometer (SLED), a new
instrument under construction for imaging spectroscopy. It is based
on the Multi-channel Subtractive Double Pass (MSDP) technique, which
combines the advantages of filters and slit spectrographs. SLED will
observe coronal structures in the forbidden lines of FeX 6374 \AA~
and FeXIV 5303 \AA, and will measure Doppler shifts up to 150 km
s$^{-1}$ at high precision (50 m s$^{-1}$) and cadence (1 Hz). It is
optimized for studies of the dynamics of fast evolving events such
as flares or Coronal Mass Ejections (CMEs), as well as for the
detection of high-frequency waves. Observations will be performed
with the coronagraph at Lomnick\'{y} \v{S}t\'{i}t Observatory (LSO),
and will also occur during total solar eclipses as SLED is a
portable instrument.
\end{abstract}


\end{frontmatter}


\section{Introduction}
\label{sec1}

Solar activity is the primary driver of space weather events which
develop over short timescales (minutes) in solar flares and CMEs.
Such events are highly dynamic and originate in the solar atmosphere
from magnetic energy stored in active regions, sunspots
\citep{Borrero2011} and coronal loops \citep{Reale2014}. Hence,
direct velocity measurements are crucial for understanding the
physics of such phenomena. In cold layers (photosphere,
chromosphere), Doppler observations are common; however, this is not
the case at coronal temperatures.

The hot corona is imaged in the extreme ultraviolet (EUV) at 15 s
maximum cadence by the Atmospheric Imaging Assembly (AIA) onboard
the Solar Dynamics Observatory (SDO), at temperatures in the range
0.6 to 10.0 MK. STEREO and the Solar and Heliospheric Observatory
(SOHO) use similar wavebands plus wide field, white light
coronagraphs, while Solar Orbiter will soon offer EUV images. In
parallel, ground-based coronagraphs provide observations at a much
higher cadence, such as the SECIS instrument \citep{Phillips2000},
working in the green line of FeXIV 5303 \AA. The corona has also
been observed during total solar eclipses in visible emission lines
\citep{Rudawy2019}, as well as in infrared (IR) lines
\citep{Habbal2011}.

Imaging instruments cannot measure plasma velocities, and for this
tunable filters or spectrographs are required. LASCO/C1 onboard SOHO
was able to scan the green (FeXIV) and red (FeX) lines with a Fabry
P\'{e}rot filter \citep{Brueckner1995}. \cite{Mierla2008} studied
velocities in the quiet corona, but it was not possible to access
the velocities of fast events such as CMEs, due to the long time
needed to scan the profiles. \cite{Singh2002} have characterized the
coronal green (FeXIV 5303 \AA) and red (FeX 6374 \AA) emission lines
using narrow slit spectroscopy. Spectra of the green line were also
recorded by \cite{Sakurai2002}. \cite{Livingston1981} and
\cite{Singh1982} introduced multi-slit spectroscopy during the total
solar eclipse of 16 February 1980 to measure velocities. The Coronal
Multi-channel Polarimeter (CoMP, \cite{Tomczyk2008}) is a
ground-based instrument working in the IR lines of FeXIII (10747
\AA~ and 10798 \AA). CoMP can record velocities (and magnetic
fields) with 15 s temporal resolution; it was used, for instance, to
study Alfv\'{e}n waves \citep{Tomczyk2007} or the dynamics of CMEs
\citep{Tomczyk2013}.

The SLED instrument described in this paper will allow the
measurement of line-of-sight (LOS) velocities in the coronal green
and red lines, in a 2D field of view (FOV) and at high cadence (1
Hz), by combining the advantages of spectroscopy and tunable
filters. Section ~\ref{sec:MSDP} explains the principles of the
imaging spectroscopy concept used by the SLED, while Section
~\ref{sec:desi} summarizes the optical design. Section
~\ref{sec:SLED} presents the main scientific goals of the instrument
and simulations of its capabilities are detailed in Section
~\ref{sec:SLEDcap}.

\section{The Multichannel Subtractive Double Pass (MSDP) technique} \label{sec:MSDP}

\begin{figure*}
\centering
\includegraphics[scale=1.0]{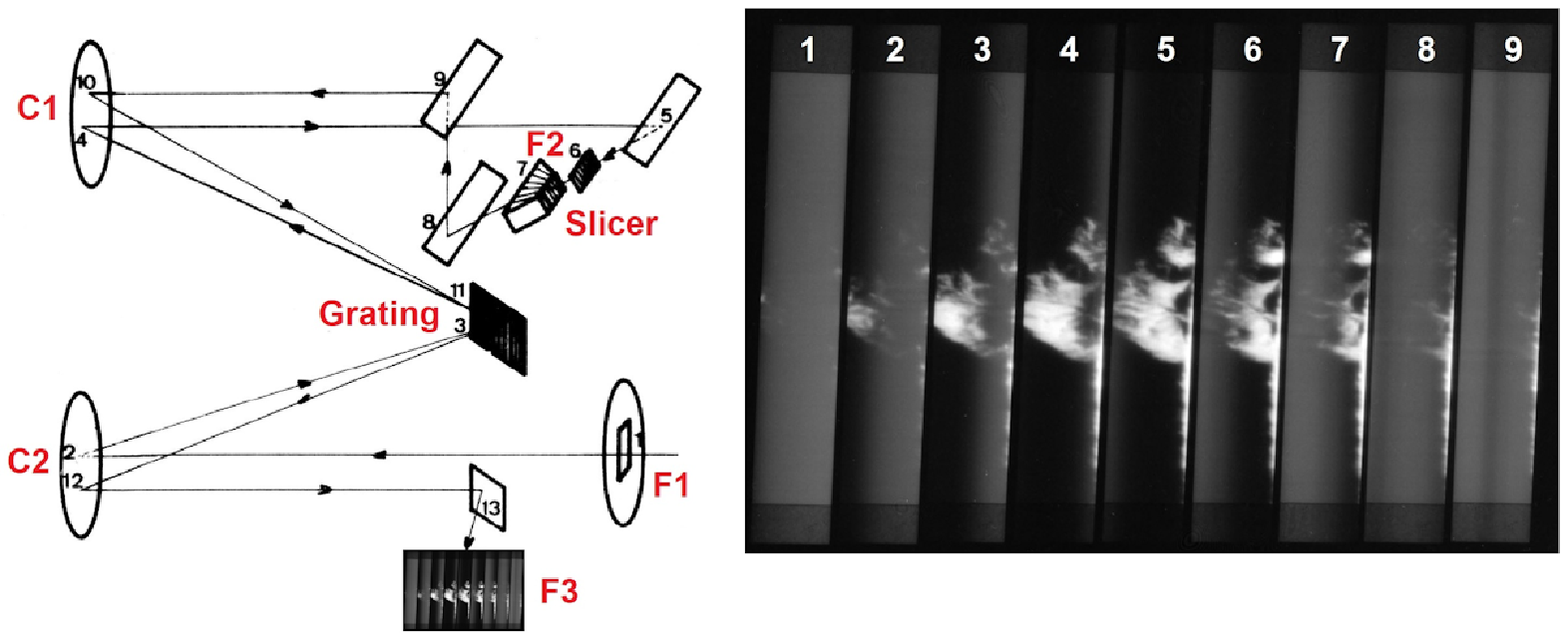}
\caption{Left: principle of MSDP spectrographs. F1 = entrance
window. F2 = spectrum. F3 = output with N spectra-images. C1 and C2
are collimators. The numbers (1 to 13) indicate the light path.
Right: an example of a prominence at the limb observed in H$\alpha$
line with N = 9 channels.} \label{dpsm}
\end{figure*}

SLED is a compact version of the MSDP imaging-spectrographs; the
principle was first described by \cite{TOUR77}, and has been
upgraded many times over the last decades on several telescopes
\citep{Mein2021}. The layout of MSDP is shown in Figure~\ref{dpsm}
with a sample set of past observations. It is an imaging double pass
spectrograph using a 2D rectangular entrance window (F1) and a
slicer located in the focal plane of the spectrum (F2), after a
first pass on the grating. The slicer selects N channels
(beam-splitting) and realigns the N channels (beam-shifting) before
the subtraction of the dispersion by the second pass on the grating.
The output of the MSDP (F3) is composed of N contiguous
spectra-images (N = 9 in Figure~\ref{dpsm}). There is a constant
wavelength step between each spectra-image, but within each, the
wavelength varies linearly along the x-direction. Data cubes (x, y,
$\lambda$) of the whole FOV are extracted from a single exposure.
Hence, the MSDP combines the advantages of filters and spectroscopy.
The 2D FOV depends on the focal length of the telescope and the size
of the window (F1), while the spectral resolution depends on the
focal length of the spectrograph and the slicer parameters. The
cadence is only limited by the photon flux and the detector speed.

The SLED/MSDP parameters are optimized for coronal lines (0.3 \AA~
resolution), high velocities (N = 24), the size of active regions
(150$''$ $\times$ 1000$''$ FOV) and time-scales involved in fast
events or high-frequency waves (1 s).

\section{The SLED optical design} \label{sec:desi}

The SLED has spectral resolution and wavelength range of
respectively 0.28 \AA~ and 6.5 \AA~ (24 $\times$ 0.28 \AA) for the
FeXIV 5303 \AA~ forbidden line. These parameters are convenient for
coronal line widths (typically 0.8 \AA~ FWHM) and large Doppler
shifts ($\pm$ 100 km s$^{-1}$ $\approx$ $\pm$ 2.0 \AA).

\begin{figure*}
\centering
\includegraphics[scale=0.13]{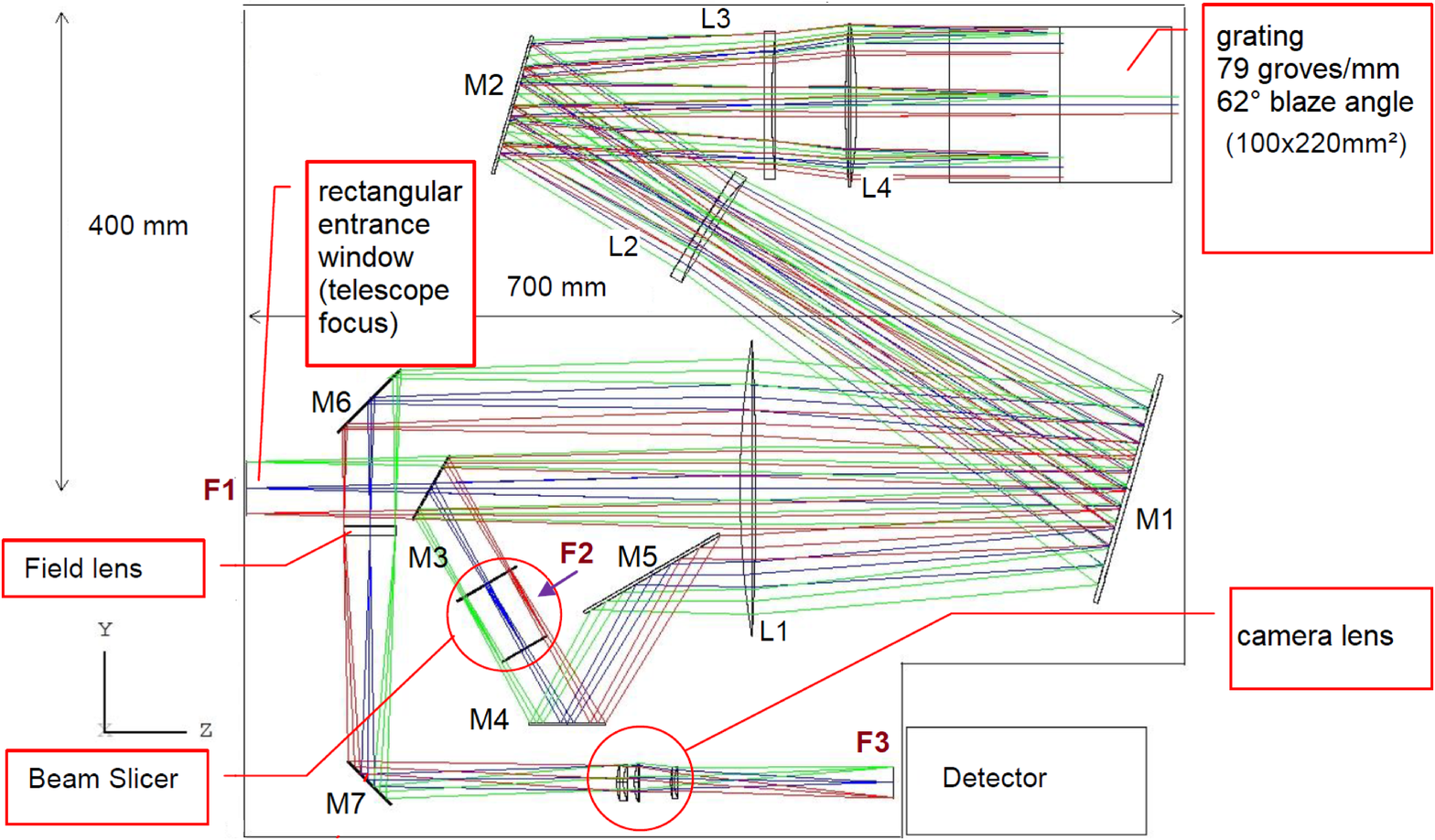}
\caption{The optical design of the SLED spectrograph. The layout is
described in detail in Section~ \ref{sec:desi}. } \label{design}
\end{figure*}

The optical path of the SLED spectrograph is schematized in
Figure~\ref{design}, where the letters L and M refer to lenses and
plane mirrors, respectively. The set of four lenses L1 to L4 acts as
a collimator. Optical components of the spectrograph are:

\begin{itemize}
  \item F1: the rectangular entrance window (4.4 mm x 31.0 mm) located at the spectrograph focus.
  \item The first pass (light dispersion): F1, L1, M1, L2, M2, L3, L4, grating,
  L4, L3, M2, L2, M1, L1, M3, slicer (F2).
  \item F2: 24 channels slicer located in the spectrum, made of 24 beam-splitting micro-mirrors and 24 associated
beam-shifting mirrors.
  \item The second pass (subtractive dispersion): slicer (F2), M4, M5, L1, M1, L2, M2,
  L3, L4, grating, L4, L3, M2, L2, M1, L1.
  \item The reduction optics to the detector: M6, field lens, M7 and camera lens (0.2 magnification).
  \item F3: the Andor Zyla camera (5.5 Mpixels sCMOS detector, 2560 $\times$ 2160 format, 6.5 micron
  square pixels) recording the 24 channel spectra-images with 2.1$''$ spatial sampling.
\end{itemize}

The optical combination of four lenses (L1, L2, L3, L4) acts as a
collimator (2.0 m equivalent focal length) with folding mirrors (M1,
M2), while the grating is 62\degree~ blazed and 79 grooves/mm ruled.
Orders are selected by filters (order 42 for FeXIV), and the
resolution is R = 19000, with dispersion D = 1.43 mm/\AA~ (for
FeXIV). The wavelength sampling (0.28 \AA) is the ratio S/D, where S
= 0.4 mm is the micro-mirror step of the slicer. On exit, the camera
lens (100 mm focal length) has a magnification factor of 0.2. More
details are given in \cite{Mein2021}.

There is a major difference between SLED/MSDP and conventional
integral field spectrographs such as the GREGOR Infrared
Spectrograph (GRIS, \cite{Vega2016}) and MuSICa for the future
European Solar Telescope \citep{Calcines2013}. Specifically, SLED
has a large variation of its spectral window coverage as a function
of the FOV position in the direction parallel to the dispersion. In
the centre of the FOV, the spectral range provided by the 24
channels is [-3.25 \AA, +3.25 \AA], corresponding to velocities up
to 160 km s$^{-1}$. Towards the FOV edges, the spectral range keeps
the same width (6.5 \AA) but is shifted from [-4.8 \AA, +1.7 \AA] to
[-1.7 \AA, +4.8 \AA] respectively from the left to the right side,
where velocities up to 75 km s$^{-1}$ can still be determined. This
3.1 \AA~ shift between both sides is the ratio W/D, where W = 4.4 mm
is the width of the 2D entrance window.

\section{The scientific goals of the SLED} \label{sec:SLED}

The SLED has two main goals owing to its capability to produce
Dopplergrams at high temporal resolution.

\subsection{Dynamic events of solar activity}

SLED will observe the dynamics of coronal flux ropes and loops in
active regions, flares, and CMEs, which drive solar-terrestrial
interactions. It will study the plasma motions (Doppler shifts)
inside hot structures (1 to 2 MK) at high cadence (1 Hz). These
measurements could be combined with EUV intensities obtained at
lower cadence by AIA onboard SDO in several lines emitted by plasma
at temperatures in the range 0.6 to 10 MK. SLED will impose new
constraints on models describing the dynamics of hot loops and their
temporal evolution. The instrument will be available before the
forthcoming solar maximum (cycle 25), when major events are expected
to occur.

\subsection{Coronal heating}

Processes involved in the coronal heating are not yet fully
understood \citep{Klimchuk2006}. The main plausible mechanisms fall
into two categories: heating by numerous and concentrated magnetic
reconnections (nano-flares), or the dissipation of Alfv\'{e}n or MHD
waves in coronal loops. Observations with CoMP \citep{Tomczyk2007}
revealed Alfv\'{e}n-type wave motions of low frequency (peaking at
3.5 mHz) around active regions at the limb. However, their energy is
too small to explain the heating, for which higher frequency waves
are likely to be involved. Owing to telemetry limitations of
satellites, a search for high-frequency waves was undertaken with
ground-based instruments in fast imagery mode through narrow-band
filters in the visible FeXIV or FeX lines. \cite{Samanta2016}
searched for high frequency oscillations (0.1 Hz) using multi-slit
spectroscopy of these two lines, observed simultaneously during the
eclipse of 11 July 2010. Several campaigns were organized with SECIS
(\cite{Rudawy2004}, \cite{Rudawy2010}, \cite{Rudawy2019}), but the
short period fluctuations were questionable in terms of waves. For
this reason, the SLED will investigate oscillations using Doppler
measurements.

\section{Simulation of the SLED capabilities} \label{sec:SLEDcap}

The core of the SLED is the 24 channel-slicer designed for F/30
beam. It is built for 0.20 m diameter/6.0 m equivalent focal length
telescopes, such as the high-altitude Lomnick\'{y} \v{S}t\'{\i}t
Observatory (LSO) coronagraph \citep{Lexa1963}. However, the SLED
will also be used for total eclipse observations, as it is a
portable spectrograph. The FOV is compatible with the typical size
of coronal active region loops ($10^{5}$ km).

\begin{figure*}
\centering
\includegraphics[scale=0.2]{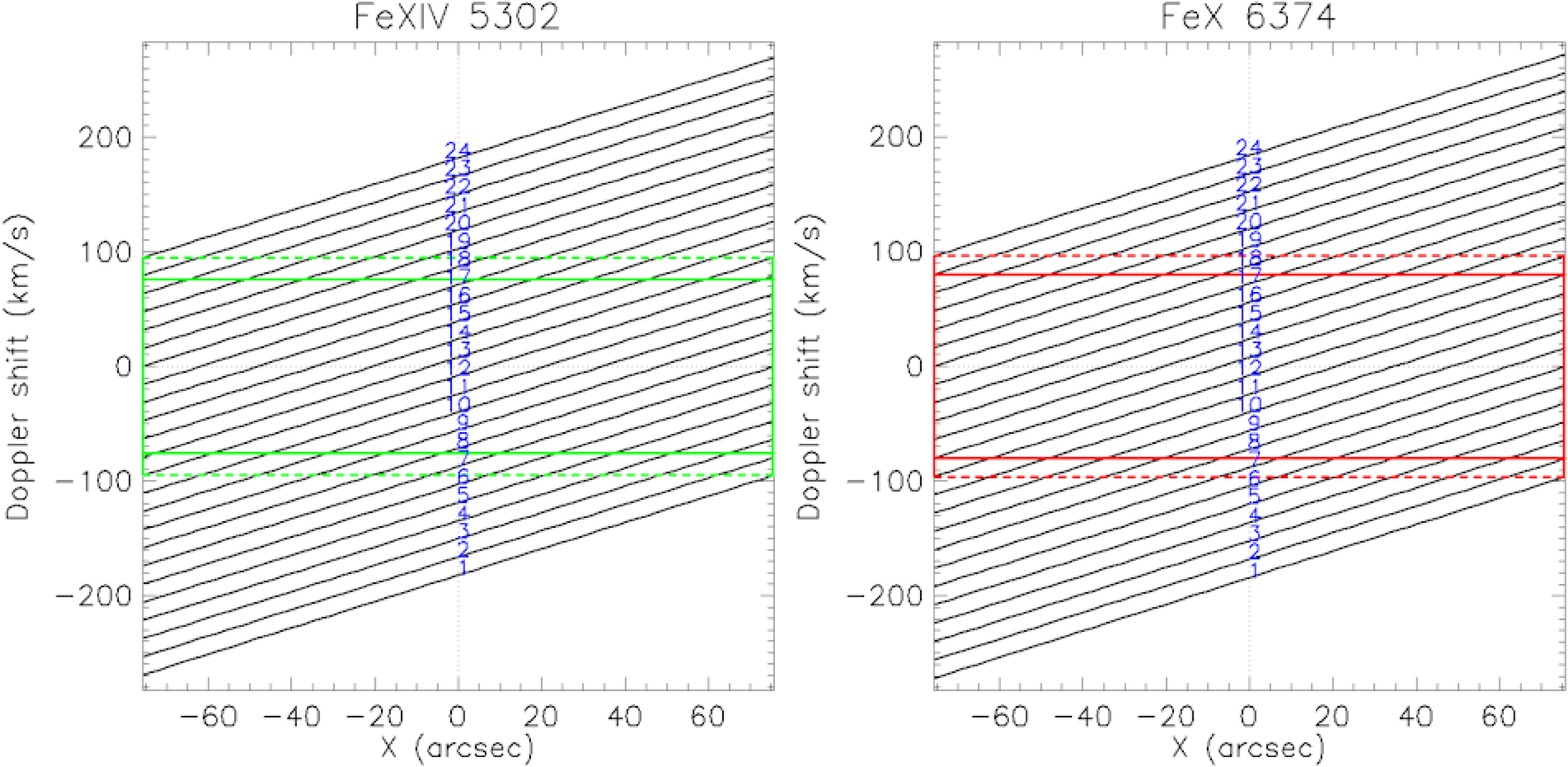}
\caption{Wavelength transmissions of the 24 channels (numbers 1 -
24) of the SLED for the FeXIV green line (left) and the FeX red line
(right). Abscissa: x direction along the 2D FOV ($''$); ordinate:
wavelength converted to LOS velocities (km s$^{-1}$). The
solid/dashed rectangles indicate the velocity range available over
the full FOV, provided by measurements respectively at the inflexion
points or peaks of the emission profiles.} \label{trans}
\end{figure*}

In Figure~\ref{trans} the wavelength response of the 24 channels in
terms of Doppler shifts (converted to velocities, positive for blue
shifts) is displayed for the coronal green and red lines. The
channels are not monochromatic, as the wavelength varies in the
x-direction of the 2D FOV. Velocities up to 75 km s$^{-1}$ can be
determined at all points in the FOV. However, velocities of the
order of the local sound speed (150 km s$^{-1}$) can be measured
near the FOV centre. For the green line, the maximum velocity is
given by the relation $\lvert v_{max} \rvert = -1.17 \lvert \times
 \rvert + 163$ km s$^{-1}$ where x is the abscissa ($''$) along the
FOV. Hence, the SLED is ideally suited to observations of highly
dynamic phenomena during solar activity.

\begin{figure*}
\centering
\includegraphics[scale=0.18]{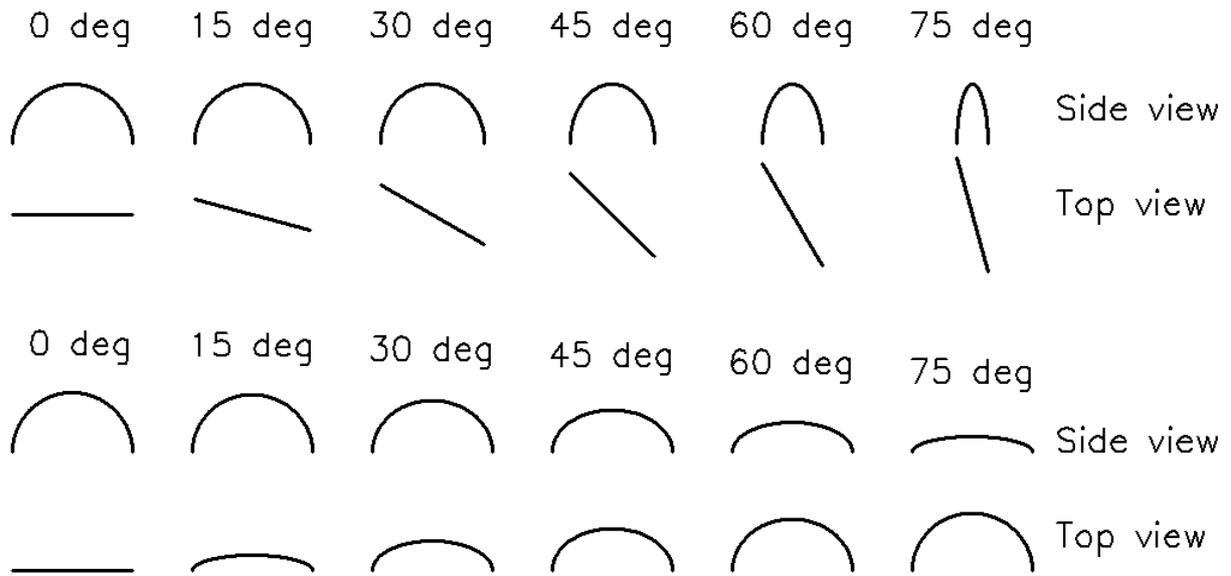}
\caption{Semi-circular magnetic flux tube seen at the limb under
various conditions and angles. Top: the plane of the loop is
orthogonal to the solar surface, but is rotated with respect to the
plane of the sky (except for 0\degree). Bottom: the plane of the
loop is inclined with respect to the local vertical (except for
0\degree).} \label{dessin}
\end{figure*}

\begin{figure*}
\centering
\includegraphics[scale=0.4]{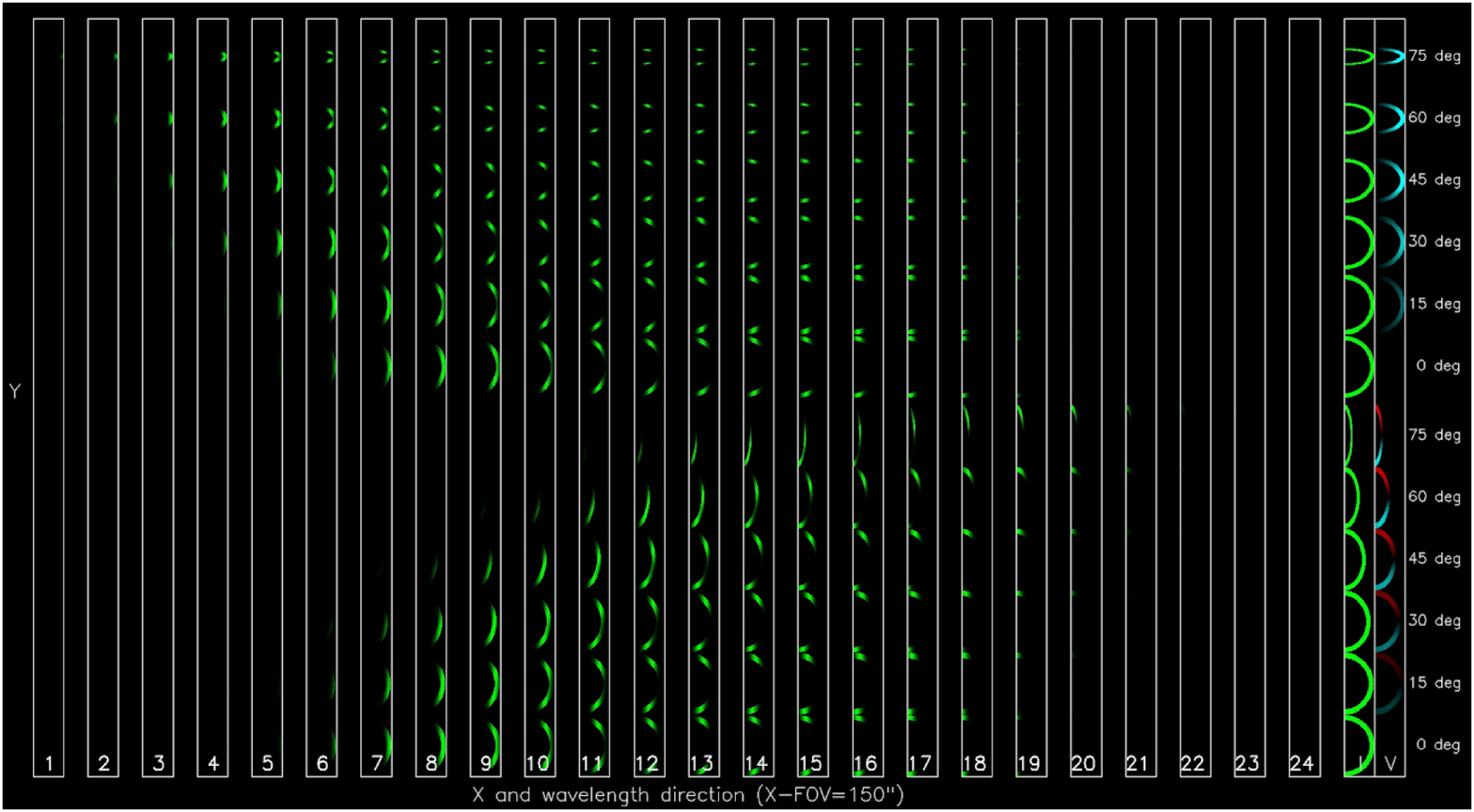}
\caption{Spectra images of the SLED in the case of a loop with
subsonic flows observed at the limb under the different viewing
conditions of Figure~\ref{dessin}. The 24 channels are shown
together with the input intensity (I) and LOS velocities (V) derived
from the siphon flow model and the various angles. Blue/red colour
are coding blue and red shifts of the model.} \label{reponse}
\end{figure*}

In order to demonstrate the capabilities of the SLED to restore LOS
velocities, we used the results of the simple model of
\cite{Cargill1980}. It is a steady-state and adiabatic siphon flow
in a semi-circular coronal magnetic loop with constant cross-section
(see Figure 6a of their paper). They suggested several possible
regimes, such as subsonic, transonic and supersonic flows with
shocks (when the local Mach number M exceeds unity). Assuming
observations at the limb, we explored different view angles of the
loop with respect to the line-of-sight, shown in
Figure~\ref{dessin}. The response of the SLED in terms of spectra
images is simulated in Figure~\ref{reponse} in the case of a
subsonic siphon flow observed in the FeXIV green line with a
Gaussian profile of 0.8 \AA~ FWHM.

\begin{figure*}
\centering
\includegraphics[scale=0.07]{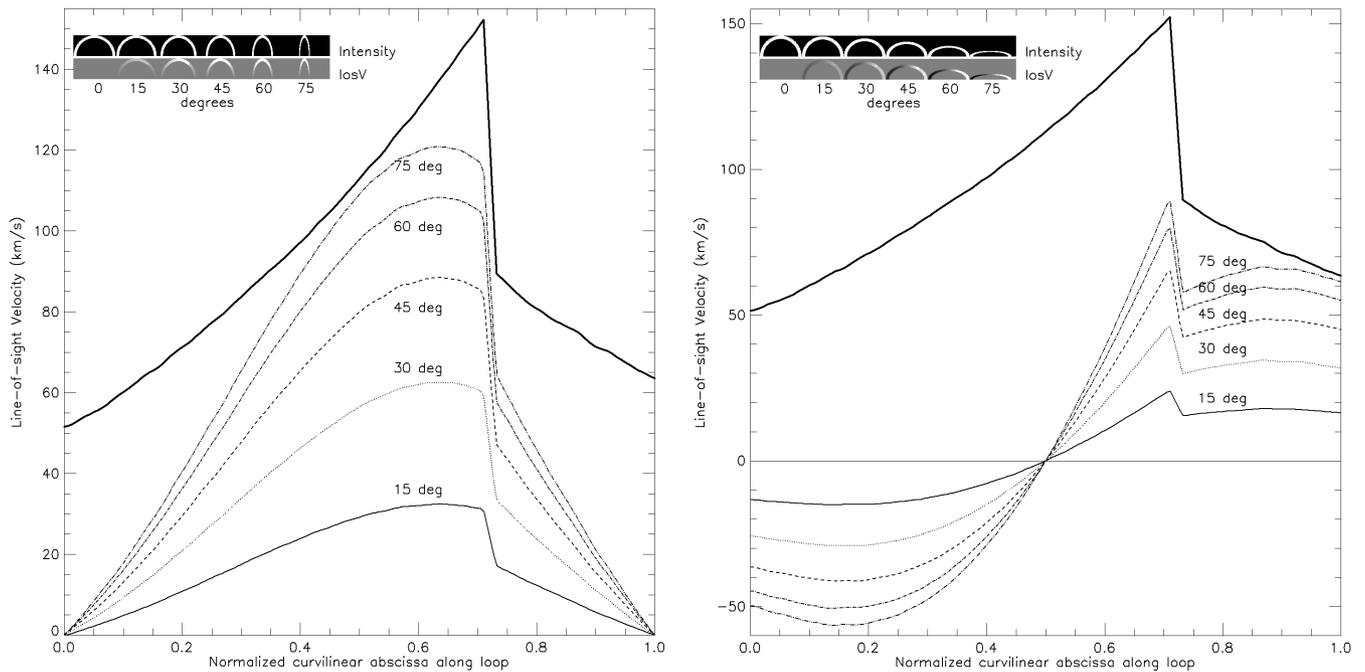}
\caption{Restoration of LOS velocities with the FOV orthogonal to
the limb. Abscissa: normalized curvilinear coordinate s along the
loop. The thick line represents the velocities of the input model,
which are tangential to the semi-circular loop (shock at s = 0.7).
The two panels correspond to the two sets of view angles of
Figure~\ref{dessin}. The thin, dotted, dashed and dot-dashed lines
are respectively for 15\degree, 30\degree, 45\degree, 60\degree~ and
75\degree~ view angles.} \label{dirx}
\end{figure*}

From the SLED spectra image, such as that in Figure~\ref{reponse},
we restored LOS velocities of subsonic and supersonic shocked flows
seen through a rectangular 2D FOV either tangential or orthogonal to
the limb. The SLED samples the line profiles with a 0.28 \AA~ step
by the transfer functions of Figure~\ref{trans}. We used a two-pass
cubic interpolation to optimize the reconstruction of line profiles.
LOS velocities were then computed using the bisector technique, in
which we detect the middle position of a chord of 0.70 \AA~ width,
corresponding to the distance between the inflexion points. This
choice minimizes errors in the velocity measurements. We found an
excellent precision of 50 m s$^{-1}$ RMS, so that the SLED
restorations of Figures~\ref{dirx} and ~\ref{diry} are
indistinguishable from the LOS velocities of the input model. Table
1 shows the RMS interpolation error for different chords: it clearly
indicates that inflexion points provide the best precision (the
slope of the profiles is maximum there). We have also simulated the
effect of the photon noise: it must be considered when the signal to
noise ratio is smaller than about 100. We estimate (from SECIS
extrapolation) that it should be 50 for the 1 Hz cadence necessary
to study high frequency waves, and above 100 for moderate frame
rates. Asymmetrical profiles can also affect the precision, for
instance in the case of velocity gradients along the line of sight.

   \begin{table}
      \caption[]{Precision of velocity measurements (IP = inflexion points).}
         \label{capab}
     $$
         \begin{array}{|c|c|}
            \hline
            \noalign{\smallskip}
            Chord~ length~ (\AA) & RMS~ error~ (km/s)  \\
            \hline
            \noalign{\smallskip}
           0.56 & 0.25 \\
           0.63 & 0.11 \\
           0.70~ (\approx IP) & 0.05 \\
           0.77~ (\approx FWHM)& 0.17 \\
           0.84~ (\approx FWHM) & 0.23 \\
            \noalign{\smallskip}
            \hline
          \end{array}
     $$
   \end{table}

\begin{figure*}
\centering
\includegraphics[scale=0.25]{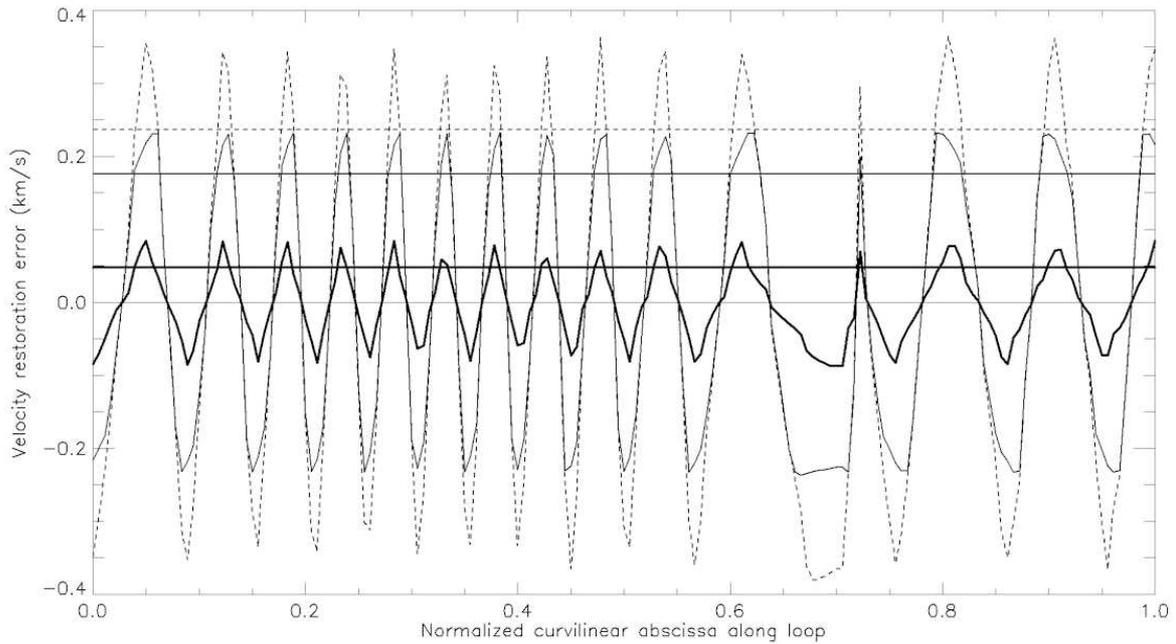}
\caption{Precision (km s$^{-1}$)  of the restoration of LOS
velocities shown in Figure~\ref{dirx} for 45\degree~ view angle.
Abscissa: normalized curvilinear coordinate s along the loop. The
thick, thin and dashed lines represent respectively the error for
chords of 0.70 \AA~ (inflexion points), 0.77 \AA~ and 0.84 \AA~
(FWHM). Associated horizontal lines give the RMS error level. The
shock is located at s = 0.7.} \label{err}
\end{figure*}

\begin{figure*}
\centering
\includegraphics[scale=0.07]{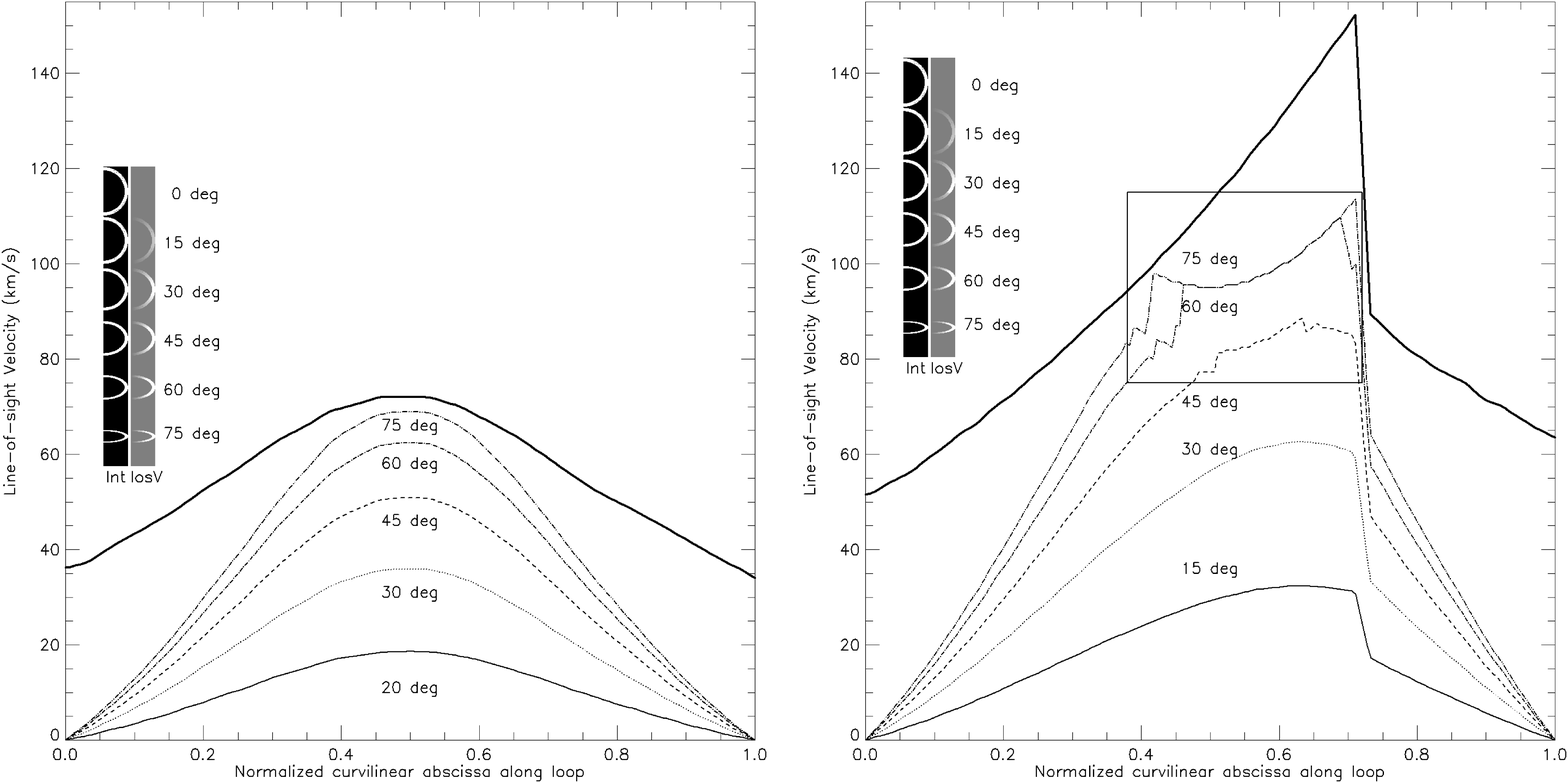}
\caption{Restoration of LOS velocities with the FOV tangential to
the limb. Abscissa: normalized curvilinear coordinate s along the
loop. The thick line represents the velocities of the input model,
which are tangential to the semi-circular loop. The thin, dotted,
dashed and dot-dashed lines are respectively for 15\degree,
30\degree, 45\degree, 60\degree~ and 75\degree~ view angles. Left:
subsonic model. Right: shocked model (shock at s = 0.7); the box
indicates a saturation phenomenon happening when the plane of the
loop is strongly rotated with respect from the plane of the sky.}
\label{diry}
\end{figure*}

Figure~\ref{dirx} shows the restoration of LOS velocities (using the
inflexion points) in the case of high-speed supersonic shocked flows
(M = 1.5) when the (long) y-direction of the rectangular FOV is
orthogonal to the limb, for the various view angles of
Figure~\ref{dessin}. The curvilinear abscissa s along the flux rope
is normalized to 1, and the shock is located at s = 0.7. SLED can
measure LOS velocities up to the sound speed when the shocked region
is close to the centre of the FOV in the x-direction. Within this
orientation, the radius of the loop cannot exceed 55000 km
(half-width of the FOV). Figure~\ref{err} displays the precision of
the velocity restoration along the loop for 45\degree~ view angle;
the best result is obtained when Doppler shifts are measured near
the inflexion points of spectral lines. Oscillations correspond to
interpolation errors occurring along the loop, because the sampling
wavelengths of the profiles vary according to the LOS velocities and
the x-abscissae in the FOV.

Alternatively, Figure~\ref{diry} shows the restoration of LOS
velocities when the (long) y-direction of the FOV is tangential to
the limb, for the various angles of Figure~\ref{dessin}. Within this
orientation, the maximum radius of the loop is 110000 km (the width
of the FOV). In the left panel of the figure, we used the subsonic
model, so that LOS velocities are easily restored everywhere in the
FOV. However, in the right panel, we used the supersonic shocked
flow model (M = 1.5), and the highest LOS velocities, now close to
the sides of the FOV, can no longer be properly recovered for all
angles. Hence, the FOV orientation has to be considered when large
radial velocities are suspected in coronal structures.

\section{Conclusions} \label{sec:Conclu}

The Solar Line Emission Dopplerometer (SLED) is a new instrument for
studies of the dynamics of coronal structures in the forbidden lines
of FeX and FeXIV. Dopplergrams providing line-of-sight velocities
will be delivered at high cadence (1 Hz) for a large FOV at the limb
(150$''$ $\times$ 1000$''$, 2.1$''$ pixel sampling). Plasma motions
can be investigated during fast-evolving events such as flares or
CMEs, and high-frequency oscillations relevant to coronal heating
will be searched. Full line profiles (0.28 \AA~ resolution) will be
available everywhere in the 2D FOV, owing to an imaging spectroscopy
technique which is faster than most tunable filters or slit
spectrographs. However, the spectral window is not the same for
different abscissae in the FOV, but has a constant width. SLED will
complement spectroscopic instruments operating mainly in IR lines,
such as CoMP or the future VELC onboard Aditya-L1 \citep{Raj2018}.
It will monitor the coronal activity of cycle 25 at Lomnick\'{y}
\v{S}t\'{\i}t Observatory, but it could also observe total solar
eclipses, such as the event of 8 April 2024 (North America).

\section{Acknowledgments}

We are indebted to the two referees for helpful comments and
suggestions. We are grateful for financial support to the Institut
National des Sciences de l'Univers (INSU/CNRS), the University of
Wroc{\l}aw, the UK Science and Technology Facilities Council (STFC),
the Leverhulme Trust via grant RPG-2019-371, and Queen's University
Belfast. J.R. acknowledges support by the Science Grant Agency
project VEGA 2/0048/20 (Slovakia).

\bibliographystyle{model5-names}
\biboptions{authoryear}
\bibliography{papier}

\newpage

\end{document}